
\magnification=1200
\vsize=8.5truein
\hsize=6truein
\baselineskip=20pt
\centerline{\bf Influence of the electric field on edge dislocations
in smectics}
\centerline{by}
\centerline{Robert Ho{\l}yst$^{1,2}$ and Patrick Oswald$^2$}
\vskip 20pt
\centerline {$^1$Institute of Physical Chemistry of the
Polish Academy of Sciences, Dept. III,}
\centerline {Kasprzaka 44/52,
01224 Warsaw, Poland}
\centerline{$^2$ Ecole Normale Superieure de Lyon, Laboratoire de Physique}
\centerline{46, All\'ee d'Italie-f-69364 Lyon Cedex 07, France}
\centerline{\bf{Abstract}}
The electric field applied perpendicularly to smectic layers breaks
the rotational
symmetry of the system. Consequently, the elastic energy associated with
distortions induced by an edge dislocation diverges
logarithmically with the size of the system. In freely
suspended smectic films the dislocations in the absence of
the electric field are located
exactly in the middle of the film. The electric field, $E$, above a certain
critical
value, $E_c$, can shift them towards the surface. This critical field
(in Gauss cgs units)
is given by the following
approximate formula
$E_c=
\sqrt{4\pi B(a\gamma/\sqrt{KB}-b)/(\epsilon_a (N-2))}$. Here, $B$ and $K$
are smectic elastic constants, $\gamma$ is the surface tension, $N$ is
 the number of
smectic layers and $\epsilon_a$ is the dielectric anisotropy.
The constant $b=0.85\pm0.07$ and $a=1.45\pm0.01$. Additionally, we assumed that
$\sqrt{K/B}=d$, where $d$ is the smectic period. This formula is valid for $
\gamma/\sqrt{KB}>2$ and $N>12$.
For smaller values of the surface tension and large N
 the linear relation between $E_c^2$ and $\gamma /\sqrt{KB}$ breaks down,
since eventually for $\gamma /\sqrt{KB}\to 1$ and $N\to\infty$,
$E_c^2(N-2)$ approaches 0.
The equilibrium location of a dislocation
in the smectic film, $h_{eq}$, equals
$N d G(x)$, where $x=E^2/E_c^2$ and $G(x)$ is a function
independent of the film thickness (for $N>12$)
and of the value of the surface tension.
\vfill\eject
\centerline{\bf I Introduction}
The influence of electric field on nematic liquid crystals has been a subject
of interest due to its application in display devices[1]. Also recently the
influence of the electric field on ferroelectric smectic C$^*$[2] liquid
crystals has gained considerable attention, since these systems can be used as
electrooptical switches[3]. The electric field can easily deform the director
field
in nematics, because of their dielectric anisotropy.
In ferroelectric smectics the electric field
applied parallel to smectic layers
can easily deform the helical polarization pattern.
Here we study the influence of electric field on
layer deformations in smectic A liquid crystals.
If a field is applied perpendicularly to the layers of smectic A with negative
dielectric anisotropy ($\epsilon_a <0$), it can destabilize their regular
arrangement.
For homeotropic samples bounded by surfaces
with strong anchoring,
the layers should undergo a transition to a periodically distorted state
(Helfrich-Hurault instability [1]). This
undulation instability has not been observed since
anchoring is never strong enough to support it. The instability is thus
preempted by other instabilities  such as anchoring transitions and nucleation
of confocal textures[4] or focal conics[5]. In this article we assume
$\epsilon_a >0$, to ensure the stability of the system
and study the influence of the field on edge dislocations.

It is well known that the electric (or magnetic) field
breaks the rotational symmetry of the system, suppresses Landau-Peierls
instability  and consequently restores the true long range order[1]
in the smectic system. The latter fact has far reaching consequences for the
defects. In ordinary solids the energy of an edge dislocation
per unit length diverges logarithmically with the size of the system. In
smectic A this energy is finite and independent of the system size. However,
 in the
electric field, the smectic should behave like a true solid and
consequently the edge dislocation energy should diverge with the system size.
This effect has not been studied so far. The question remains if the electric
field
effects in finite systems can be easily measured with moderate fields
accessible in
laboratories.

In freely suspended smectic liquid crystal films edge dislocations are
stabilized in
the middle of the film[6,7]. This effect is understandable since surface
elastic
energy is larger than the bulk elastic energy and thus edge dislocations are
repelled from the limiting free surfaces (smectic-air interfaces)[8].
However, we can expect
that in a sufficiently strong electric field (above a certain
critical field) the equilibrium location of the
edge dislocation may shift towards the surface. The rest of this paper is
devoted
to the study of this effect.

The paper is organized as follows. In section II we discuss the deformations
induced by an edge dislocation in a bulk system subjected to an electric field.
In section III we do the  same calculations for a finite system
bounded by two free surfaces. The equation for the equilibrium location
of an edge dislocation within the film and a critical field are also
derived in this section. The results
are discussed in section IV.

\centerline{\bf II Bulk system}

The elastic energy density of a weakly deformed smectic A liquid crystal
in the electric field, $E$,
along the z-axis (perpendicular
to smectic layers),
is given by the
following equation[1]:
$$f_b={1\over{2}}\left(B\left({{\partial u({\bf r})}\over
{\partial z}}\right)^2+{{\epsilon_a E^2}\over{8\pi}}\left\vert
\nabla_{\bot} u(\bf r)\right\vert^2+
K(\triangle_\bot
u({\bf r}))^2\right).\eqno(2.1)$$
The vertical displacement in the system at point ${\bf r}$
is $u({\bf r})$, $K$ and $B$ are the curvature and compressional
elastic constants,
$\epsilon_a$ is the dielectric anisotropy. We assume that $\epsilon_a >0$.
The total elastic energy,
$$F_b=\int d{\bf r} f_b\eqno(2.2)$$
minimized with respect to $u$ leads to the following equation for the
distortions in the system:
$$-{{\partial^2 u}\over{\partial z^2}}-
f_e\triangle_\bot u+\lambda^2\triangle_\bot^2 u
=0.\eqno(2.3)$$
Here $f_e=\epsilon_a E^2/(4\pi B)$ is a dimensionless constant and
$\lambda =\sqrt{K/B}$ a microscopic length.

The elementary edge dislocation (of Burgers vector,b,
 equal to the layer thickness, d)
located at $z=h$ and $x=0$ is characterized by the
following condition:
$$u(x,z=h^{\pm})=\cases{0, &if $x\le 0$;\cr {\rm sgn}(z-h)d/2,&if
$x>0$.\cr}\eqno(2.4)$$
The displacement $u$ is a multivalued function in the $z=h$ plane.
The equilibrium solution that satisfies Eq(2.3) and condition (2.4) and thus
minimizes the elastic energy in the presence of the dislocation is as follows:
$$u_b(x,z)={\rm sgn}(z-h)\left({d\over{4\pi}}
\int dq \exp{(-a(q)\vert z-h\vert)}{{\exp{(iq x)}}\over{i(q-i0^+)}}
\right),\eqno(2.5)$$
where
$$ a(q)=\sqrt{\lambda^2 q^4+f_e q^2}\eqno(2.6)$$
The elastic energy of the deformations induced by the dislocation can
be conveniently written in the following form:
$$F_b=F_c+{1\over 2}B\int dx{{\partial u_b(x,z)}\over{\partial z}}
\left(u_b(x,z=h^{-})-u_b(x,z=h^{+})\right),\eqno(2.7)$$
where the derivative is evaluated at
$z=h$. Here  $F_c$ is the core energy.
Inserting $u_b$ into Eq(2.7) gives:
$$F_b=F_c+{{B d^2}\over{4\pi}}\int_{2\pi /L}^{2\pi/r_c} d q
{{a(q)}\over{q^2}}\eqno(2.8)$$
Here $r_c$ is the core size and $L$ is the size of the system in the $x$
direction
(in calculations one introduces the
cutoff in the $q$ integral[9]).
It is assumed that the
system is infinite in the $z$ direction. For $E=0$ we find the elastic energy
($L\gg r_c$),
$$F_b=F_c+\sqrt{KB} d^2/2r_c.\eqno(2.9)$$
independent of $L$. However in the electric field , $E\ne 0$,
the energy diverges logarithmically with the size $L$ of the system.
We find the following approximate form of $F_b$ in this case:
$$F_b=F_c+{{B d^2\sqrt{f_e}}\over{4\pi}}\ln{\left({{\sqrt{f_e} L}\over
{2\pi\lambda}}\right)}
+\sqrt{KB} d^2/2r_c\left(1-{{\sqrt{f_e} r_c}\over{2\pi\lambda}}\right),
\eqno(2.10)$$
In fact it also means that edge dislocations
should interact with long range potential even when placed in the same plane,
when in the absence of the electric field their interactions vanish[9,10].

The same observations are valid
if we set $L\to\infty$, but assume finite size $D$ in the z direction.
Then, the energy
diverges logaritmically with $D$. However, in this case, we have to specify
boundary conditions at the surfaces bounding the smectic. This is done in the
next section.

\centerline{\bf III Dislocations in films}

In a symmetric freely suspended smectic film of thickness $D$ subjected to an
electric field $E$ the elastic energy for each of the  bounding surfaces
located at
$z=\pm D/2$ reads as follows[6,8]:
$$F_s={1\over 2}\int d{\bf r}_\bot\left(\gamma\vert\nabla_\bot u_s
\vert^2+{{\epsilon_a E^2 d}\over{8\pi}}\left\vert\nabla_\bot u_s\right\vert^2+
K d(\triangle_\bot
u_s)^2\right).\eqno(3.1)$$
Here $u_s=u(x,z=\pm D/2)$ is the displacement of surface
layers.
The fact that the curvature  energy  and the
electric energy are not only included in the bulk term
but also in the surface term, follows
from the comparison of the distortion energy in the discrete and
continuous representation.
In the discrete representation the layers have indices $i=0,1,2...$
and $u(x,z)=u_i(x)$. Then the curvature and electric terms appear naturally
for every layer including the surface one.
In the discrete representation we sum over $i$ instead
of integrating over $z$ and consequently
for each layer, including the surface one, the electric term and the curvature
term
are multiplied by the smectic period $d$. Although for freely suspended films
these terms can be neglected in comparison to the surface term[6-8],
nonetheless
in systems characterized by small surface tension they are by no means
negligible
[11]. That is why we shall keep them, although we notice that their
influence on
the calculated quantities is rather small for the systems considered
in this paper. In particular
for large $D$ and large surface tension $\gamma$ one could use
the following simple and well known surface energy:
$$F_s={1\over 2}\int d{\bf r}_\bot\gamma\vert\nabla u_s
\vert^2.\eqno(3.2)$$
without considerable change of results.
At the end we shall discuss the
differences between the results obtained from Eq(3.1)
and those from Eq(3.2).

The minimization of $F_b+F_s$ (using Eq(3.1))
leads to Eq(2.3) in the bulk and the following
boundary conditions:
$$-B{{\partial u}\over{\partial z}}-\left(\gamma+{{\epsilon_a E^2d}\over
{4\pi}}\right)
{{\partial^2 u(x,z=-D/2)}\over {\partial x^2}}+K d
{{\partial^4 u(x,z=-D/2)}\over {\partial x^4}}=0\eqno(3.3)$$
at $z=-D/2$ and
$$B{{\partial u}\over{\partial z}}-\left(\gamma+{{\epsilon_a E^2d}
\over{4\pi}}\right)
{{\partial^2 u(x,z=D/2)}\over {\partial x^2}}+K d
{{\partial^4 u(x,z=D/2)}\over {\partial x^4}}=0\eqno(3.4)$$
at $z=D/2$.
The solution satisfying Eqs(2.3,3.3,3.4) and condition (2.4) is
as follows:
$$u(x,z)=u_b(x,z)+u_p(x,z)\eqno(3.5)$$
where $u_b$ is given by Eq(2.5) and
$$u_p(x,z)={d\over{4\pi}}\int dq{{\exp{(iq x)}}\over{iq}}
\left(\exp{(-a(q) z)}f_1(q)+\exp{(a(q) z)}f_2(q)\right).
\eqno(3.6)$$
The function $a(q)$ is defined by Eq(2.6),
$$f_1(q)={{a_1(q)(a_2(q)e^{a(q)(D+h)}+
a_1(q)e^{-a(q)h})}\over
{a_2(q)^2e^{2 a(q) D}-a_1(q)^2}}
,\eqno(3.7)$$
$$f_2(q)=-{{a_1(q)(a_2(q)e^{a(q)(D-h)}+
a_1(q)e^{a(q)h})}\over
{a_2(q)^2e^{2 a(q) D}-a_1(q)^2}}
,\eqno(3.8)$$
and
$$a_1(q)=\left(
\gamma+{{\epsilon_a E^2d}\over{4\pi}}\right) q^2+Kd q^4-B a(q),\eqno(3.9)$$
$$a_2(q)=\left(
\gamma+{{\epsilon_a E^2d}\over{4\pi}}\right) q^2+Kd q^4+B a(q).\eqno(3.10)$$
The energy associated with distortions induced in the smectic film
of thickness $D$ by the dislocation located at distance $h$ from the
center of the film is given by:
$$F(h)=F_b+{1\over 2}B\int dx{{\partial u_p(x,z)}\over{\partial z}}
\left(u_b(x,z=h^{-})-u_b(x,z=h^{+})\right),\eqno(3.11)$$
which for finite $D$ is finite, but which in the large $D$ limit diverges as
$\ln{(D)}$. The equilibrium location of the dislocation is determined by the
minimum of $F(h)$ as a function of $h$
i.e. by the solution of the following equation:
$$\int_0^{\infty} d q \left({{a(q)}\over q}\right)^2 {{a_1(q)a_2(q) e^{a(q)
D}\sinh{(
2a(q)h)}}\over{a_2(q)^2e^{2 a(q) D}-a_1(q)^2}}=0.\eqno(3.12)$$
For $\gamma >\sqrt{KB}$ and small electric field $E$ the dislocation is located
exactly in the middle of the film at $h=0$. However above a certain critical
value of the electric field, $E_c$, the minimum at $h=0$ becomes unstable and
the dislocation moves out of the center of the film. The condition
which determines the critical field is given by the following equation:
$$\int_0^{\infty} d q \left({{a(q)}\over q}\right)^2 {{a_1(q)a_2(q) e^{a(q) D}
a(q)}\over{a_2(q)^2e^{2 a(q) D}-a_1(q)^2}}=0,\eqno(3.13)$$
which corresponds to the vanishing of the second derivative of
$F(h)$ with respect to $h$ at $h=0$.
The solution of Eq(3.13) for the value of the critical field and of Eq(3.12)
for
the equilibrium location of the dislocation in the film $h_{eq}$ are discussed
in the next section.

\centerline{\bf IV Results and discussion}

The electric term has the same formal structure as the surface tension term.
Consequently, since the location of the dislocation in the film depends on the
competition between surface and bulk terms,
we can expect that when the electric term multiplied by the size of the film
is roughly proportional to the surface term, the dislocation will move
from the middle of the film towards the surface. This condition can be simply
written as follows:
$${{\epsilon_a E_c^2}\over{4\pi}} D\sim\gamma\eqno(4.1)$$
Indeed, we find the following relation between the critical field $E_c$ and the
surface tension (assuming that  the microscopic length, $\lambda$, is equal
to $d$):
$${{\epsilon_a E_c^2}\over{4\pi B}} (N-2)\approx a{{\gamma}\over{\sqrt{KB}}}-b,
\eqno(4.2)$$
Here, $N=D/d$ is the number of smectic layers; $a\approx 1.45\pm 0.01$ and
$b\approx 0.85\pm 0.07$
are two constants. This approximate formula works very well for
$\gamma >2\sqrt{KB}$ and $N>12$. The difference between this approximate
formula
and numerical solution of Eq(3.13) is within few percents.
In fact for typical experimental samples
of thickness 1$\mu$m ($N\approx 300$),
with $\gamma\approx 5\sqrt{KB}$, this difference is smaller than one percent.

For large N and $\gamma\to\sqrt{KB}$,
$E_c^2 (N-2)$ approaches 0 and the linear relation (4.2) breaks down.
In Fig.1 the critical field, $E_c^2 (N-2)\epsilon_a/(4\pi B)$, is plotted
versus $\gamma/\sqrt{KB}$ for $N=12$ and $N=102$ for
$\gamma/\sqrt{KB}\le 2$.
As $\gamma$ increases the differences
between the two cases decreases. Practically for $N\to\infty$,
$E_c^2 (N-2)\epsilon_a/(4\pi B)$, should follow the same curve as for $N=102$
except
in the very small region close to $\gamma/\sqrt{KB}=1$.

We can estimate the value of the critical field for typical liquid crystal
sample
characterized by large $\epsilon_a$ e.g. cyanobiphenyl. For
$\epsilon_a=10$[12],
$\gamma /\sqrt{KB}=5$ [13] (in some lamellar systems of polymers
it can be as large as 30
[14]), $N=300$ and $B=10^7$ dyn/cm$^2$ we find
$$E_c\approx 15{\rm V}/{\rm \mu m}\eqno(4.3)$$
Such fields are easily obtained in typical experiments [15] and thus the
influence of the electric field on an edge dislocation inside the film should
be visible in experiments. Without the field two
dislocations located in the
center of the film could not cross easily. On the other hand, in the electric
field,
 two dislocations
located close to different surfaces can cross easily, which should be visible
under the microscope.

Using the simplified surface elastic energy Eq(3.2) (instead of (3.1))
we find a formula similar to
(4.2), where instead of $N-2$ there is $N$.
However for $\gamma/\sqrt{KB}\le 2$, the differences are much larger. In
particular for $\gamma/\sqrt{KB}\to 1$, $E_c¬2 (N-2)\to 0$
independently of the film thickness.

Finally, we calculate the equilibrium location of the dislocation, $h_{eq}$, in
the film.
For $E<E_c$ we have  $h_{eq}=0$, while for $E\ge E_c$ we find
the following relation between
the electric field and $h_{eq}$:
$$h_{eq}=\pm N d G\left({{E^2}\over {E_c^2}}\right).\eqno(4.4)$$
The function $G$, shown in Fig.2,
is independent of the film thickness and the surface tension.
It rises steeply for $E\geq E_c$, which means that
for $E$ even sligthly above the critical field dislocations move far away
from the middle of the film.
For large electric fields the dislocations approach the surface i.e.
$h_{eq}\to (D/2-d)$.  Since the function $G$ is very flat for
large values of $E$, thus for practical reasons, it should be very hard to
shift the
dislocation exactly at the surface by using the electric field only.

Another question of practical interest is to know whether
the dislocation can easily reach its new equilibrium position
after the electric field
is applied, since there is in the system a kinetic barrier, known
as Peierls-Nabarro stress, opposing the
motion of the dislocation in the direction perpendicular to the layers.
Thus,in order to move the dislocation,
 the glide force,$F_g$, acting on the dislocation must be comparable or larger
than the Peierls-Nabarro force, $F_{P.N.}$,
pinning the dislocation in its Peierls valley.
The latter has been calculated by Lejcek[16]:
$$F_{P.N.}={3\over 4}B b\sqrt{{{\pi\lambda}\over{r_c}}}
\exp{\left(-{{2\pi r_c}\over b}
\right)}.\eqno(4.5)$$
 It strongly depends on the size of the dislocation core $r_c$, which
at low temperature is of the order of $d$, but should be larger than d
near a nematic-smectic phase transition.
The glide force in our case equals
$$F_g=\left\vert{{\partial F(h)}\over{\partial h}}\right\vert\eqno(4.6)$$
This force vanishes for $h=0$ and $h=h_{eq}$ and passes through the maximal
value, $F_g^{max}$,
in between. For $E\sim 3 E_c$ and $N=102$ we find
$$F_g^{max}\approx 0.01 {{B d}\over{2\pi}}\eqno(4.7)$$
{}From Eq(4.6) and (4.7) we find that for $r_c=d$
$$F_{P.N.}\gg F_g^{max},\eqno(4.8)$$
but for $r_c\ge 2d$
$$F_{P.N.}\ll F_g^{max}\eqno(4.9)$$
Consequently, we can expect that the phenomena studied here can be
easily observed near a second order (or weakly first order)
nematic-smectic phase transition when
the core of the dislocation is probably dissociated (Eq(4.9)).
Otherwise, dislocations should be strongly pinned in the middle of the
film (Eq(4.8)).

Although our results are not applicable for the case of negative anisotropy
we can see that in such systems the dislocations would be favored by the
electric field, applied perpendicular to layers.
It opens an interesting problem of the nucleation of
dislocations in such systems subjected to the electric field. It is known
for example that for
the
ferroelectric liquid crystals the film can assume a cone configuration with the
stable pore running through the film[17]. Similarly in our system
the dislocation loops could be stabilized by the electric field.

\centerline{\bf Acknowledgements}

The research has been supported in part
by the KBN grant 303 020 07 and 30219004. RH acknowlegdes with
appreciation the support from the CNRS and the hospitality of the Ecole Normale
Superieure de Lyon.
\vfill\eject

\centerline{\bf References}
\item{[1]} P.G. de Gennes and J.Prost, {\it The Physics of Liquid Crystals},
Oxford Science Publications, 1993.
\item{[2]} R.B.Meyer, L.Liebert and P.Keller, J.Phys.Lett (Paris) {\bf 36}, 69
(1975).
\item{[3]} N.A.Clark and S.T. Lagerwall, Appl.Phys.Lett. {\bf 36}, 899 (1980);
M.A.Handschy, N.Clark and S.T.Lagerwall, Phys.Rev.Lett. {\bf 51}, 471  (1983).
\item{[4]} J.B.Fournier, Phys.Rev.Lett. {\bf 70}, 1445 (1993).
\item{[5]} Z.Li and O.D.Lavrentovich, Phys.Rev.Lett. {\bf 73}, 280 (1994).
\item{[6]} L.Lejcek and P.Oswald, J.Phys.II {\bf 1}, 931 (1991);
R.Ho\l yst, Phys.Rev.Lett. {\bf 72}, 4097 (1994).
\item{[7]} P.Piera\`nski et al, Physica A {\bf 194}, 364 (1993).
\item{[8]} for a review on this topic see R.Ho\l yst and P.Oswald,
Int.J.Mod.Phys.B
(in press) (1995).
\item{[9]} M.Kl\'eman, {\it Points, Lines and Walls}, Wiley, 1983.
\item{[10]} S.Chandrasekhar and G.S.Ranganath, Adv. Phys. {\bf 35}, 507 (1986).
\item{[11]}R.Ho\l yst, Macromol. Theory and Simul. {\bf 3}, 817 (1994).
\item{[12]} H.Kelker and R.Hatz, {\it Handbook of Liquid Crystals},
Verlag Chemie, 1980.
\item{[13]} D.J.Tweet,R.Ho\l yst,B.D.Swanson,H.Stragier
and L.B.Sorensen,Phys.Rev.Lett. {\bf 65}, 2157 (1990).
\item{[14]} M.S.Turner, M.Maaloum, D.Ausserr\'e, J.-F. Joanny and M.Kunz,
J.Phys. II
(France), 689 (1994).
\item{[15]} P.E.Cladis, H.R.Brand and P.L.Finn, Phys.Rev. A {\bf 28}, 512
(1983).
\item{[16]} L.Lejcek, Czech. J Phys. {\bf B 32}, 767 (1982).
\item{[17]} J.Prost and L.Lejcek, Phys.Rev. A, {\bf 40}, 2672 (1989).
\vfill\eject
\centerline{\bf Figure Captions}

\item{Fig.1} $\epsilon_a E_c^2(N-2)/(4\pi B)$ versus $\gamma/\sqrt{KB}$
for $N=12$ (dashed line) and $N=102$ (solid line). Here $E_c$ is the critical
field,
$\gamma$ -- surface tension and $N$ -- number of smectic layers.
\item{Fig.2} The equilibrium location of the dislocation, $h_{eq}/D$, versus
$E^2/E_c^2$. Here $E$ is the electric field, $E_c$ -- critical field, and
$D$ -- thickness of the film. $h_{eq}/D$=0 corresponds to the middle of the
film and
$h_{eq}/D$=0.5 to the surface.
\vfill\eject\end